\documentclass[aps,prl,twocolumn,groupedaddress]{revtex4}

\usepackage{graphicx} 
\usepackage{dcolumn}
\usepackage{bm}

\begin{document}
\title{Reply to Comment on the paper ``Pairing mechanism of high-temperature superconductivity:
Experimental constraints" } 
\author{Guo-meng Zhao$^{1,2,*}$} 
\affiliation{$^{1}$Department of Physics and Astronomy, 
California State University, Los Angeles, CA 90032, USA~\\
$^{2}$Department of Physics, Faculty of Science, Ningbo
University, Ningbo, P. R. China}

\begin{abstract}

In our recent paper entitled "Pairing mechanism of high-temperature superconductivity:
Experimental constraints" (to be published in  Physica Scripta), we 
review some crucial experiments that place strong constraints on the microscopic pairing 
mechanism of high-temperature superconductivity in cuprates. In
particular, we show that phonons rather than spin-fluctuation play 
a predominant role in the microscopic pairing 
mechanism.  We further show that
the intrinsic pairing symmetry in the bulk is not $d$-wave, but
extended $s$-wave (having eight line nodes) in hole-doped cuprates and 
nodeless $s$-wave in electron-doped cuprates.  
In contrast, the author of the Comment  
(to be published in  Physica Scripta) argues that our conclusions are unconvincing and even misleading. In response to the criticisms in
the Comment, we further show that our conclusions are well supported by experiments and his 
criticisms are lack of scientific ground.

\end{abstract}
\maketitle 

In our recent paper \cite{Zhao}, we 
review some crucial experiments that place strong constraints on the microscopic pairing 
mechanism of high-temperature superconductivity in cuprates. In
particular, we show that phonons rather than spin-fluctuation play 
a predominant role in the microscopic pairing 
mechanism.  We further show that
the intrinsic pairing symmetry in the bulk of cuprates is not $d$-wave, but
extended $s$-wave (having eight line nodes) in hole-doped cuprates and 
nodeless $s$-wave in electron-doped cuprates.  
However, Plakida \cite{Pla} has raised strong criticisms on these conclusions 
based on an oversimplified polaronic model and some experimental
results that have been misinterpreted. Below we will show that our conclusions are well 
supported by experiments and the criticisms raised in the Comment are lack of scientific ground.

In the Comment \cite{Pla}, the author first considers the oxygen-isotope
effect (OIE) on $T_{c}$ by taking into account the observed
oxygen-isotope effect on the in-plane effective supercarrier mass. He 
has used the weak-coupling BCS-like formula [Eq.~(2) in the Comment] to calculate the doping
dependence of the OIE on
$T_{c}$ in La$_{2-x}$Sr$_{x}$CuO$_{4-y}$ (LSCO) on the assumption that the
electron-phonon coupling constant $\lambda_{ep}$ has the same OIE on the in-plane effective 
supercarrier mass. Actually, Eq.~(2)  used in the Comment  is incorrect.  The correct $T_{c}$ formula  in the (bi)polaron theory has the polaronic half bandwidth in front of the exponent  rather than the phonon energy \cite{aleiso}. When the correct expression is applied, the theory describes well the doping dependence of the isotope exponents in many cuprate superconductors \cite{aleiso}. Also, the tunneling experiments
\cite{Gonnelli,Shim} have consistently 
shown that $\lambda_{ep}$ is larger than 2.5 for optimally doped Bi$_{2}$Sr$_{2}$CaCu$_{2}$O$_{8+y}$ (BSCCO).
Therefore, the weak-coupling BCS-like formula does not hold in cuprates. A
strong-coupling formula has been used to consistently explain the negligible OIE on
$T_{c}$ and substantial OIE on the in-plane effective supercarrier mass
in nearly optimally doped BSCCO \cite{Zhaoisotope}. In the underdoped regime, superconductivity
should be better described by the Bose-Einstein condensation of local pairs in the strong coupling limit
\cite{alemot}. In this case, $T_{c}$ is essentially proportional to
$n_{s}/m^{**}_{ab}$ (where $n_{s}$ is the supercarrier density and $m^{**}_{ab}$ 
is the in-plane effective supercarrier mass), in agreement with the
well-known Uemura plot \cite{Uemura}. This implies that the OIE on $T_{c}$ is
essentially proportional to the OIE on $m^{**}_{ab}$. This scenario
can naturally explain why the magnitudes of the exponents for the OIE on
both $T_{c}$ and $m^{**}_{ab}$ increase with the decrease of doping
and are even larger than 0.5 in deeply underdoped samples
(e.g., La$_{1.94}$Sr$_{0.06}$CuO$_{4-y}$) \cite{ZhaoJPCM98,Zhao}.   Therefore,  the 
claim that the polaronic effects cannot explain 
the doping dependence of the OIE on $T_{c}$ is lack of scientific ground. 

Then the author of the Comment \cite{Pla} attempts to explain the doping
dependence of the OIE on $T_{c}$ in terms of his own theory based on
the $t-J$ model. According to his model \cite{Pla}, $\Delta T_{c}/T_{c}$ =
$(1/\lambda) \Delta J/J$, where $\lambda$ = 0.2$-$0.3. Since the
OIE on $J$ was found to be  $-$0.9$\%$ for undoped
YBa$_{2}$Cu$_{3}$O$_{6}$ (Ref.~\cite{ZhaoAF07}), the predicted $\Delta T_{c}/T_{c}$ for the
optimally doped YBa$_{2}$Cu$_{3}$O$_{6.94}$ should be $-(2.7-4.5)\%$, 
in disagreement with the measured  value of $-$0.27$\%$
(Ref.~\cite{ZhaoYBCO95}). In the deeply
underdoped regime, $T_{c}$ $\propto$ $1/m^{**}_{ab}$ and $1/m^{**}_{ab}$
$\propto$ $J$ within the $t-J$ model, so $\Delta T_{c}/T_{c}$ =
$\Delta J/J$ = $-(0.6-0.9)\%$, in disagreement with the measured 
values of $-(5-12)\%$ (Ref.~\cite{ZhaoJPCM98}). Therefore, the $t-J$ model
cannot explain the doping dependence of the OIE on $T_{c}$. Further,
the author of the Comment \cite{Pla} argues that since the site-selective OIE shows that the OIE
on $T_{c}$ mainly contributes from the planar oxygen, the apical
oxygen is not important for the pairing. This argument is in parallel 
with the statement that since the optimally 
doped YBa$_{2}$Cu$_{3}$O$_{6.94}$ has a negligible OIE on $T_{c}$, phonons are not important 
to the pairing mechanism. In fact, the site-selective OIE experiment
\cite{ZhaoPr} shows that 
the apical oxygen contributes about 40$\%$ of
the OIE on $m^{**}_{ab}$ in YBa$_{2}$Cu$_{3}$O$_{6.94}$. This suggests that the apical oxygen is
important to the pairing, in agreement with our argument based on the 
bulk-sensitive x-ray-absorption experiment \cite{Merz} on (Y$_{1-x}$Ca$_{x}$)Ba$_{2}$Cu$_{3}$O$_{7-y}$.

The author of the Comment also criticizes the conclusion about strong 
coupling to multiple phonon modes revealed by both tunneling
\cite{Gonnelli,Shim,Zhaopairing07,Boz08,ZhaoPRL09,ZhaoSM07} and angle-resolved photoemission spectra (ARPES)
\cite{Zhou04,ZhaoSM07}.
There are several important facts about the strong coupling features
revealed by both tunneling and ARPES data. First, the energies of strong coupling features 
match very well with those of the phonon modes revealed by the neutron
\cite{Zhaopairing07} and
Raman \cite{Boz08} data. Second, the energies of strong coupling features
revealed by tunneling spectra match very well with those of strong coupling features
revealed by ARPES \cite{ZhaoSM07}. Third, the energies of strong coupling features in 
different cuprate systems such as LSCO, YBCO, and BSCCO are
very similar \cite{Zhou04,Zhaopairing07,ZhaoSM07} and nearly independent of doping \cite{Mak}. Such excellent
consistencies unambiguously demonstrate that these strong-coupling features 
are intrinsic and arise from strong electron-phonon interactions. 
Furthermore, a detailed review of tunneling, ARPES, and optical
experiments has recently been given by Maksimov {\em et al.}
\cite{Mak}.
These authors provide consistent evidence for strong coupling to multiple
phonon modes from the tunneling, ARPES, and optical results. 

Concerning the spin-fluctuation pairing,  we have shown that the magnetic
resonance mode revealed by neutron experiments plays a minor role in high-temperature 
superconductivity \cite{Zhao,ZhaoPRL09}. 
One might argues that since the magnetic resonance mode is only a small fraction of the spin-excitation
spectrum \cite{Kiv}, its contribution to the electron pairing should be
insignificant while the coupling to the whole spin-excitation spectrum
may still play an important role in the $d$-wave pairing.  A recent theoretical calculation
\cite{Sca} has shown that strong coupling to the whole spin-excitation spectrum of underdoped
YBa$_{2}$Cu$_{3}$O$_{6.6}$ (measured by neutron scattering) can lead
to $d$-wave high-temperature superconductivity with $T_{c}$ = 174 K
and the coupling constant $\lambda_{d}$ = 1.39. It is important to
note that the authors of Ref.~\cite{Sca} have used a large renormalized coupling 
strength $\bar{U}$ (1.59 eV), which is too large compared with that estimated 
from several other independent experiments (see a 
recent review article \cite{Mak}). These experiments consistently show
that \cite{Mak,Kiv} $\bar{U}$ $<$ 0.17 eV. This implies that $\lambda_{d}$
$<$ 0.0115 (since $\lambda_{d}$ $\propto$ $\bar{U}^{2}$). Such a small
coupling constant cannot lead to high-temperature superconductivity.
Further experimental evidence for no magnetic pairing mechanism is
that $T_{c}$ is very insensitive to the magnetic spectral weight, as
clearly demonstrated from the neutron data \cite{Bou} of slightly underdoped
YBa$_{2}$Cu$_{3}$O$_{6.92}$ and slightly overdoped
YBa$_{2}$Cu$_{3}$O$_{6.97}$. The two compounds have almost the same
$T_{c}$ (91-93 K), but the magnetic spectral weight for YBa$_{2}$Cu$_{3}$O$_{6.97}$
is at least three times smaller than that for YBa$_{2}$Cu$_{3}$O$_{6.92}$.
As pointed out by  Maksimov {\em et al.} \cite{Mak}, the little dependence of $T_{c}$ on the magnetic spectral weight is
incompatible with the magnetic pairing mechanism.  On the theoretical ground, 
recent variational Monte Carlo  simulations \cite{imada}, which are based on an advanced sign-problem-free Gaussian-Basis Monte Carlo algorithm,  have shown  that the simplest Hubbard model, advocated by Plakida and  some other authors,  does not account for high-temperature superconductivity.

Another important issue is the intrinsic pairing symmetry in the
bulk of superconducting cuprates. Because nearly all the surface and phase-sensitive
experiments for both electron- and hole-doped cuprates provide
clear evidence for $d$-wave order-parameter (OP) symmetry \cite{Review}, the $d$-wave
pairing symmetry has become an indisputable fact to most researchers
in the field. However, it is important to note that these surface and phase-sensitive experiments 
based on planar Josephson tunneling  
are probing the OP symmetry at surfaces and interfaces,
which were found to be underdoped \cite{Bet,Mann}. Since the majority of charge
carriers are oxygen-hole bipolarons in the underdoped regime
\cite{Zhao} and the OP symmetry of the Bose-Einstein
condensate of 
the oxygen-hole bipolarons is $d$-wave \cite{AlexSM98}, the
phase-sensitive experiments just probe the $d$-wave OP symmetry of the
dominant component.  Since the OP symmetry of the Bose-Einstein
condensate has nothing to do with the pairing symmetry, the
phase-sensitive experiments do not probe the pairing symmetry
associated with the pairing interaction. In order to probe the
intrinsic pairing
symmetry, bulk-sensitive data should be obtained from significantly overdoped samples where
the dominant charge carriers are Fermi-liquid-like and the
superconducting transition is mean-field-like \cite{Zhao}. Based on the quantitative analyses of
many bulk-sensitive experiments (in addition to some bulk- and  phase-sensitive experiments such as
nonmagnetic pair-breaking effects), we have concluded that the intrinsic pairing symmetry in 
the bulk of cuprates is not $d$-wave, but
extended $s$-wave (having eight line nodes) in hole-doped cuprates
\cite{ZhaoSM01} and 
nodeless $s$-wave in electron-doped cuprates
\cite{ZhaoPRBSM10,ZhaoJPCMSM10}.  

\begin{figure}[htb]
	 \includegraphics[height=6.2cm]{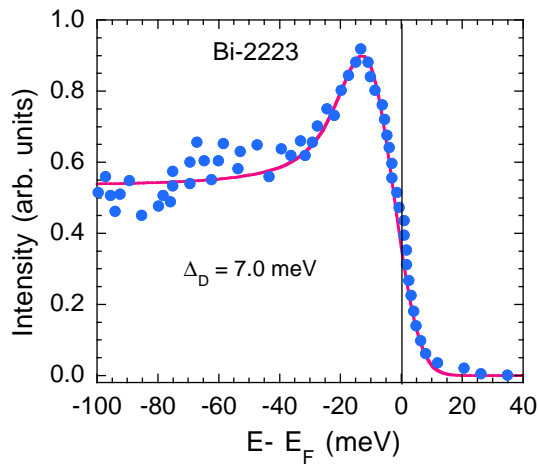}
	\caption[~]{Photoemission spectrum along the
diagonal direction for a nearly optimally doped 
Bi$_{2}$Sr$_{2}$Ca$_{2}$Cu$_{3}$O$_{10+y}$ (Bi-2223). The
spectrum was taken with an energy resolution of 10 meV (Ref.~\cite{Feng}). 
The solid line is the curve calculated with the fitting method of Ref.~\cite{Ding}
and the following parameters: $\Delta_{D}$ = 7.0 meV and $\Gamma$ = 9.0 meV. }
\end{figure}

The author of the Comment does not believe the intrinsic pairing symmetry 
inferred from the bulk and phase-sensitive nonmagnetic 
pair-breaking effects. He argues against the extended $s$-wave gap
symmetry using the surface-sensitive ARPES and Fourier transform scanning tunneling spectroscopy
(FT-STM) experiments.  
Even surface-sensitive ARPES data 
of nearly optimally doped BSCCO can be better explained in terms of 
an extended $s$-wave gap \cite{ZhaoSM01,ZhaoPM,ZhaoSM07}.  The gap along the diagonal direction
$\Delta_{D}$ is small ($\leq$ 7 meV) for 
nearly optimally doped  BSCCO \cite{ZhaoPM,ZhaoSM07} but becomes larger in heavily overdoped BSCCO
\cite{Vob,ZhaoSM01}.  
Furthermore, both surface-sensitive ARPES and 
FT-STM data of a nearly optimally overdoped BSCCO can also be well explained
in terms of an extended $s$-wave gap with $\Delta_{D}$ $\simeq$ 4 meV 
(Ref.~\cite{ZhaoPM}). The significant uncerntainty in extracting the gap size from the ARPES data 
in a slightly overdoped and two underdoped BSCCO crystals \cite{Lee07} does not allow one to make 
distinction between a $d$-wave gap and an
extended $s$-wave gap with $\Delta_{D}$ $\leq$ 4 meV. In fact, no ARPES data along the diagonal
direction were given in this study \cite{Lee07}, which makes it harder to draw a 
definitive conclusion about the gap symmetry from the ARPES data.

\begin{figure}[htb]
	 \includegraphics[height=12cm]{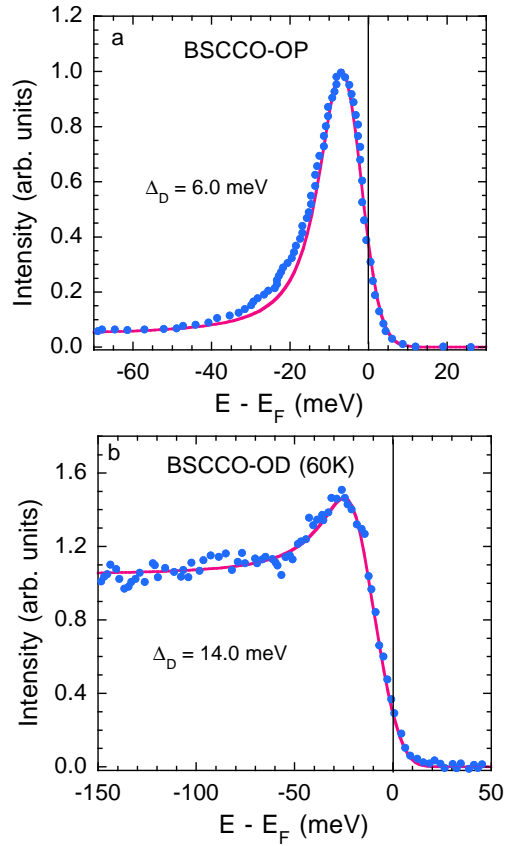}
	\caption[~]{a) Photoemission spectrum along the
diagonal direction for a nearly optimally doped 
BSCCO. The
spectrum was taken with an energy resolution of about 6 meV (Ref.~\cite{Dan}). 
b) Photoemission spectrum along the
diagonal direction for a heavily overdoped doped 
BSCCO with $T_{c}$ = 60 K. The
spectrum was taken with an energy resolution of about 10 meV (Ref.~\cite{Vob}).
The solid lines are the curves calculated with
the following parameters: $\Delta_{D}$ = 6.0 meV and $\Gamma$ = 6.0 meV 
for the nearly optimally doped BSCCO; $\Delta_{D}$ = 14 meV and $\Gamma$ = 18 meV 
for the heavily overdoped BSCCO.  }
\end{figure}

In order to further prove our extended $s$-wave pairing symmetry in
hole-doped cuprates, we determine 
the $\Delta_{D}$ value from some high-resolution ARPES data.  The fitting
method used in Ref.~\cite{Ding} for extracting the gap size from ARPES
data should be the most reliable since the gap sizes extracted from
this method match precisely with those independently determined from
the FT-STM data \cite{ZhaoPM}. Fig.~1 shows photoemission spectrum along the
diagonal direction for a nearly optimally doped Bi$_{2}$Sr$_{2}$Ca$_{2}$Cu$_{3}$O$_{10+y}$ (Bi-2223). The
spectrum was taken with an energy resolution of 10 meV (Ref.~\cite{Feng}). 
The solid line is the curve calculated with the method of Ref.~\cite{Ding}
and the following parameters: $\Delta_{D}$ = 7.0 meV and $\Gamma$ = 9.0 meV
(where $\Gamma$ is the electron life-time broadening parameter). It is
apparent that the gap size along the diagonal direction is not zero.

In Fig.~2, we show photoemission spectra along the
diagonal direction  for a nearly optimally doped BSCCO
(Fig.~1a) and a heavily overdoped BSCCO (Fig.~1b). The spectra for the nearly optimally doped and heavily overdoped BSCCO crystals were taken with  energy
resolutions of about 6 meV and 10 meV, respectively \cite{Vob,Dan}. The solid lines are the curves calculated with
the following parameters: $\Delta_{D}$ = 6.0 meV and $\Gamma$ = 6.0 meV 
for the nearly optimally doped BSCCO; $\Delta_{D}$ = 14 meV and $\Gamma$ = 18 meV 
for the heavily overdoped BSCCO.  It is striking that $\Delta_{D}$
increases from the increase of doping, in agreement with the earlier
ARPES result \cite{Kelley} and break-junction tunneling experiments \cite{ZhaoSM01}.

In summary, our conclusions of the phonon-mediated pairing mechanism and 
$s$-wave pairing symmetry in cuprates are well supported by experiments.  The 
criticisms raised in the Comment \cite{Pla} are lack of scientific ground although the author of the Comment agrees that there exist polaronic charge carriers and electron-phonon coupling is strong in cuprates.

~\\
$^{*}$ gzhao2@calstatela.edu

\bibliographystyle{prsty}

\end{document}